\documentclass[journal,twoside,web]{ieeecolor}
\usepackage{jsen}
\usepackage{cite}
\usepackage{amsmath,amssymb,amsfonts}
\usepackage{algorithmic}
\usepackage{graphicx}
\usepackage{textcomp}
\usepackage{wrapfig}

\def\BibTeX{{\rm B\kern-.05em{\sc i\kern-.025em b}\kern-.08em
    T\kern-.1667em\lower.7ex\hbox{E}\kern-.125emX}}
\definecolor{abstractbg}{rgb}{0.89804,0.94510,0.83137}
\begin{document}

\title{Absorption and Phase-Contrast Microtomography Using Direct X-Ray Detection With Commercial Off-The-Shelf CMOS Sensors}

\author{
Damián L. Corzi, 
José Lipovetzky, \IEEEmembership{Senior Member, IEEE},
Fabricio Alcalde Bessia,
Germán Mato,
Andrés Cicuttin,
María L. Crespo,
Martín Pérez,
Mariano Gómez Berisso
\thanks{D. L. Corzi (damian.corzi@ib.edu.ar), J. Lipovetzky (lipo@ib.edu.ar, corresponding author), M. Pérez, and M. Gómez Berisso are with Comisión Nacional de Energía Atómica (CNEA), Bariloche, Argentina.}
\thanks{D. L. Corzi, J. Lipovetzky, F. Alcalde Bessia, G. Mato, M. Pérez, and M. Gómez Berisso are with Instituto Balseiro, Universidad Nacional de Cuyo, Bariloche, Argentina.}
\thanks{J. Lipovetzky, F. Alcalde Bessia, G. Mato, M. Pérez, and M. Gómez Berisso are with Consejo Nacional de Investigaciones Científicas y Técnicas (CONICET), Argentina.}
\thanks{A. Cicuttin is with the Trieste Section of the Istituto Nazionale di Fisica Nucleare (INFN), Trieste, Italy.}
\thanks{M. L. Crespo is with the Multidisciplinary Laboratory, International Centre for Theoretical Physics (ICTP), Trieste, 34100, Italy.}
\thanks{M. Pérez was supported by the ICTP Associates Program (2023–2028).}
}

\IEEEtitleabstractindextext{%
\fcolorbox{abstractbg}{abstractbg}{%
\begin{minipage}{\textwidth}%
\begin{wrapfigure}[14]{r}{3in}%
\includegraphics[width=3in]{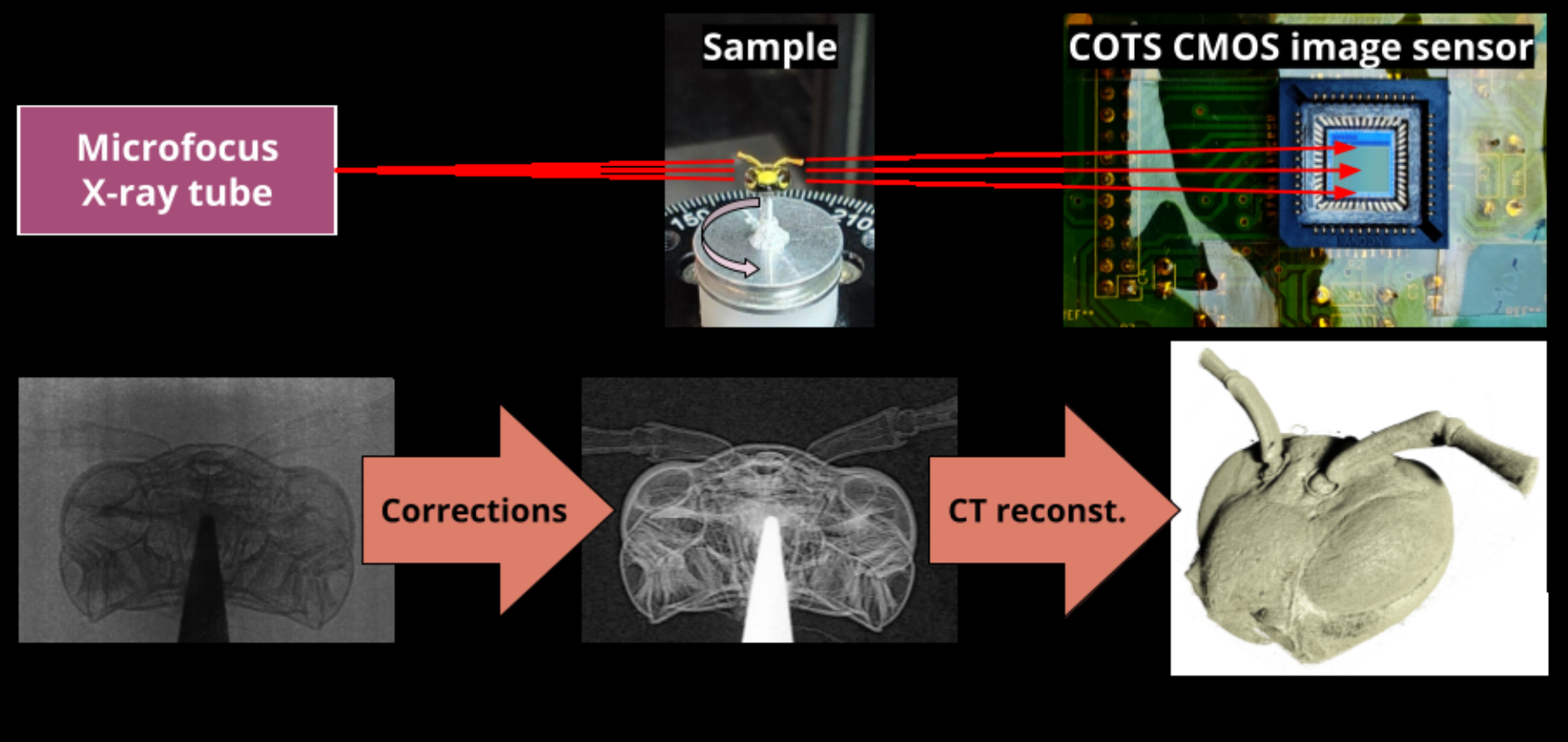}%
\end{wrapfigure}%
\begin{abstract}
This work presents high-resolution X-ray microtomography based on commercial off-the-shelf (COTS) CMOS image sensors as direct detectors. The system uses a microfocus X-ray source in cone-beam geometry, enabling both absorption-contrast imaging, thanks to the sensor's intrinsic resolution, and phase-contrast imaging, by adjusting the object-to-sensor distance. A dynamic flat-field correction algorithm mitigates radiation-induced sensor degradation during long acquisitions. The setup provides voxel sizes from $3.9\,\mu m$ to $5.2\,\mu m$, with an effective spatial resolution (FWHM) as fine as $9.6\,\mu m$ in magnified phase-contrast mode ($1.33\times$) and $21\,\mu m$ in non-magnified absorption-contrast mode. Phase contrast enhances visualization of soft tissue boundaries that are difficult to visualize using absorption-contrast imaging in the same experimental configuration.
\end{abstract}

\end{minipage}}}

\maketitle

\section{Introduction}

Commercial off-the-shelf (COTS) complementary metal-oxide-semiconductor (CMOS) image sensors (CIS)---originally designed for capturing visible light in consumer electronics---are an attractive option for high-resolution X-ray imaging as direct-radiation detectors, thanks to their small pixel pitch, high integration density, and low cost \cite{corzi-jsss, corzi2024enhanced, alcalde2018x}. Moreover, this small pixel pitch allows the observation of edge-enhancement fringes when using a microfocus X-ray source, enabling propagation-based phase-contrast imaging of low-attenuation samples with a simpler setup than other phase-contrast techniques \cite{MartinFase, Riedel2023}.

In this work, we take advantage of these features to generate high-resolution, three-dimensional tomographic reconstructions from X-ray images obtained with these sensors. We present CIS-based phase-contrast tomography, which improves on conventional absorption-contrast tomography by enhancing the edges of radiographed objects, making sample features visible even when attenuation is negligible, as with small or weakly absorbing samples.

This is because absorption contrast and phase contrast depend differently on the atomic number ($Z$). Considering the dual wave-particle nature of light, photon-matter interaction can be described using the complex refractive index $n=1-\delta+i\beta$, where $\beta$ is associated with material absorption and $\delta$ with the dispersive component \cite{stohr2023nature}. Although both coefficients decrease significantly with photon energy---except near absorption edges---$\beta$ remains smaller than $\delta$ in the low-energy X-ray range, giving techniques that exploit dispersive phenomena an advantage: enhanced image contrast with reduced energy deposition, and potentially lower radiation dose and sample damage \cite{stohr2023nature}.

Obtaining phase-contrast images requires several geometrical considerations. The interference pattern projected on the sensor must be larger than the pixel pitch; otherwise, its spatial variations are averaged within individual pixels, reducing fringe visibility. Observing this pattern also requires lateral coherence, obtained using microfocus X-ray sources with few-micrometer focal spots \cite{Gayer:13,Begani_Provinciali_s20226469}. Since the fringe size is proportional to the square root of the object-to-sensor distance \cite{Gayer:13, MartinFase}, phase contrast requires moving the object away from the detector. This benefits geometric magnification but also enlarges the penumbra---an effect kept small thanks to the use of microfocus sources.


This work demonstrates that a strategic combination of acquisition and correction techniques---dynamic flat-field correction, radiation-hardening compensation, and geometry-driven phase-contrast imaging---can turn a sensor not designed for X-ray detection into a platform capable of phase-contrast tomographic reconstruction, achieved here, to the best of our knowledge, for the first time with a COTS CMOS image sensor. The remainder of the paper describes the experimental methods and materials, presents and discusses the results, and closes with the conclusions.

\subsection{Experimental setup}

\begin{figure}[!t]
\centering\includegraphics[width=\columnwidth]{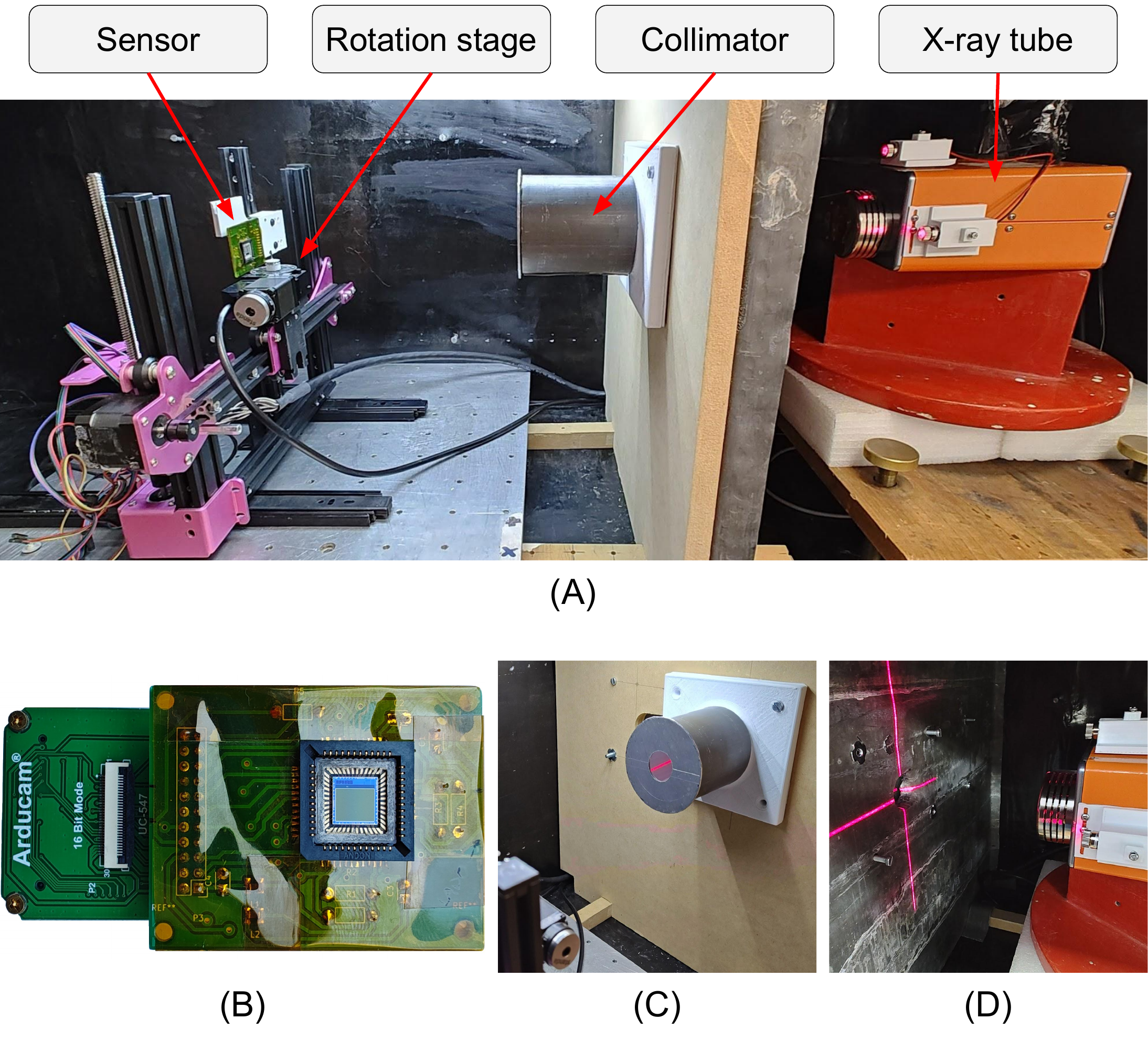}
\caption{Experimental setup used for tomographic acquisition. (A) The light-tight, shielded cabinet, divided into two compartments: the sensor and sample stage on the left, and the X-ray source on the right. (B) The sensor. (C) The internal shielding partition, which reduces scattered photons reaching the sensor. (D) The X-ray source. A detailed view of the sample holder is presented below in Figs.~\ref{Fig:chip_en_setup}~and~\ref{Fig:abeja_en_setup}.} \label{Fig:setup}
\end{figure}

The experimental setup for tomography is shown in Fig.~\ref{Fig:setup}. A constant source-to-sensor distance of 760\,mm was kept throughout all experiments, moving the object along the system's axis to exploit both geometric magnification and phase contrast. All components were arranged horizontally for better mechanical stability.

The entire irradiation setup was placed inside a closed, light-tight cabinet lined internally with at least $1\,\text{mm}$ of lead for radiation protection. The cabinet was divided into two compartments by a lead wall with a window acting as a collimator, minimizing scattered photons reaching the sensor and significantly improving the Signal-to-Noise Ratio (SNR) of the images.

The X-ray source used was a tungsten-anode Microbox Integrated X-ray Source, with a $0.254\,\text{mm}$ thick beryllium window, manufactured by Micro X-ray Inc. It has a $5\,\mu\text{m}$ focal spot when operated at power levels below $7.5\,\text{W}$, as in the phase-contrast images of this work, and supports continuous operation in the 10 to 90\,kV range, with a maximum power of 15\,W.

The sensor was an onsemi (formerly Micron) MT9M001. This CIS consists of a $1280\times1024$ pixel array with a pixel size of $5.2\times5.2\,\mu\text{m}^2$, providing a total active area of $6.66\times5.32\,\text{mm}^2$. The small pixel size allows the visualization of interference fringes in a similar geometry as was previously reported in \cite{MartinFase}.

The sensor was read and configured using an acquisition platform based on the Arducam UC-425 board and the UC-547 adapter \cite{Arducam}. The complete image acquisition system is presented in Fig.~\ref{Fig:setup}(B), showing the sensor without the transparent cover included in its original package. Removing this cover is essential to eliminate its attenuation, scattering, and fluorescence effects. The readout platform was mounted on an aluminum plate, facilitating sensor alignment and supporting a lead sheet to shield the readout electronics from Total Ionizing Dose (TID) effects.

Due to the small dimensions of the radiographed objects, they were carefully adhered to the tip of a needle mounted on an aluminum SEM pin stub holder. This holder was positioned on a Standa 8MR184(E)-11-22 rotation stage with an angular resolution of $0.015\,^{\circ}$. The stage was mounted on a motorized two-axis translation system, which travels along rails to allow easy adjustment of the object-to-sensor distance. Both the rails and the readout platform were mounted on an aluminum optical table to ensure precise projection alignment. To reduce the presence of vibrations in the acquisition system, the optical table was placed on a low-density foam mat.

To limit the sensor temperature, a cooling device based on a Tec1-12706 Peltier cell was developed to inject cold air into the cabinet. The cold air enters the sensor compartment, passes through the collimator, reaches the X-ray tube compartment, finally exiting through an opening there and dissipating the heat generated by the system. An external computer controlled the X-ray tube, rotation stage, and image acquisition.

To maximize the number of photons collected and reduce shot noise, the exposure time was configured to $1\,\text{s}$, the maximum allowed by the sensor. Since the photon flux in this specific configuration typically yields limited statistics for a single 1-second exposure, multiple frames were captured and averaged to reduce shot noise. The CIS analog gain was set to its maximum value. Automatic correction features such as white balance and defective pixel correction, typically employed in standard photography, were disabled to avoid artifacts in the images \cite{corzi2024enhanced}.

Without magnification, the field of view (FOV) of the system corresponds to the sensor size, and scales inversely with the geometrical magnification. 

\subsection{Real-Time Noise Regulation}

Because exposure time and photon flux are limited, reducing shot noise requires averaging several one-second frames. Since photon detection in CMOS sensors follows Poisson statistics, averaging improves the SNR as the square root of the number of frames, improving overall image quality.

However, tomography requires acquiring many projections, which increases TID received by the sensor over time. This leads to sensor degradation, manifested as an increase in local dark current \cite{bessia2018displacement}. While post-processing can correct the mean (deterministic) component of this increase, it cannot compensate for its random fluctuations (noise), which grow with TID as well.

To maintain the SNR of each projection above a minimum value throughout the tomography, we found it necessary to increase the number of one-second frames acquired at each angular step, compensating for the TID-induced degradation with a higher number of averaged frames. The estimation of the number of frames has to be performed online during acquisition. In our case, we used the upper 50-row band across the entire sensor width, unshaded by the object. The SNR is estimated as the ratio between the mean signal in this region and its square root, and the number of averaged frames was increased by one whenever the SNR was below the target threshold for two consecutive measurements.

\begin{figure}[!t]
\centering\includegraphics[width=\columnwidth]{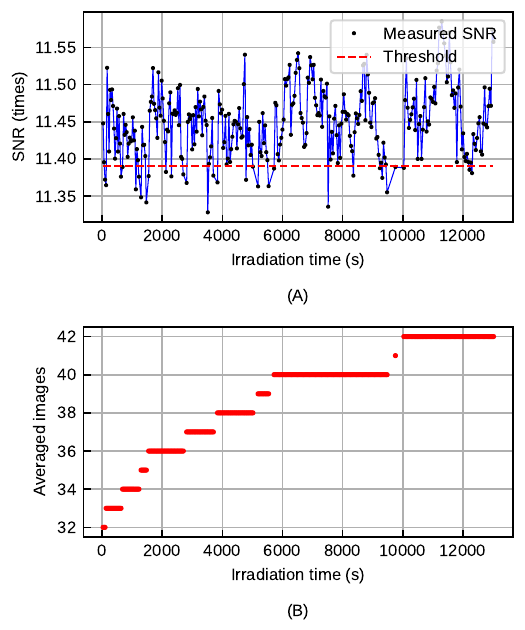}
\caption{Dynamic adjustment of the number of averaged frames to compensate for TID-induced sensor degradation during tomography. (A) SNR obtained after averaging all images acquired at each angular step. (B) Number of averaged frames per angular step, which increases from 32 to 42 as the acquisition progresses.}
\label{fig:snr_tomography}
\end{figure}

The application of this strategy is depicted in Fig.~\ref{fig:snr_tomography}, where (A) shows the SNR obtained after averaging all the images in the same angular step during the tomography presented in Section~\ref{sec:chip} of this work. The number of averaged images at each step is presented below in (B). As the tomography is being acquired, the number of frames required to comply with the target SNR increases from 32 to 42 per angular step. The target SNR threshold was determined empirically based on previous attempts, selecting the minimum value required to ensure adequate visual image quality and reliable feature extraction in the final reconstruction.

Since the number of averaged frames determines the acquisition duration at each angular position, this analysis allows us to extend the sensor's lifespan by minimizing acquisition time at the start of the operation and increasing the number of averaged frames only when necessary. As a result, this strategy avoids the need for explicit noise weighting during image reconstruction, as is done in \cite{zeng2013filtered}.

\subsection{Defective pixel correction and interpolated flat-field corrections}

After acquisition, images are corrected for defective pixels and flat-field non-uniformities---the latter compensating for local pixel gain variations and the TID-induced increase in average dark current---to minimize artifacts in tomographic reconstruction \cite{orhan2020micro, shaw2014cone}.

Defective pixel correction is applied in two stages. First, hot pixels (bright in the average of several dark frames) and dead pixels (dark in the average of irradiated frames) are identified. Then, each defective pixel is replaced by interpolating its neighbors. The same method can also mitigate saturated pixels. Although most cameras have built-in defective pixel corrections, these must be disabled since they erase single-pixel events, and the correction must instead be performed on averages of images to avoid this problem.

Flat-field correction is used to mitigate non-uniformities in local pixel dark current and response, sometimes referred to as ghosting, and is based on the acquisition of additional reference images obtained with the X-ray source turned off, denoted as Dark Frames ($DF$), and others with the sensor directly irradiated in the absence of a sample, denoted as Flat Frames ($FF$) \cite{corzi2024enhanced}.

However, during the acquisition of a large image set, a TID-induced increase in the average dark current implies that the correction parameters must be periodically updated to track this drift \cite{goiffon2015radiation}. Although dark current increase can be non-linear with dose, exhibiting time-dependent annealing, we experimentally found that in our detector, for the doses involved in short irradiation steps, it could be approximated as linear with small dose steps. For the signal span, the difference between flat and dark images, it is sufficient to assume a linear dependence along the whole irradiation. 

Assuming this linear behavior, an \textit{ad-hoc} dynamic interpolated version of the flat-field correction can be applied, as described in Eq.~\ref{Ecu:corr}. For this approach, a full set of flat-field images is acquired at both the start and the end of the irradiation process (moving the object from the beam path), and dark images are acquired every ten angular steps (turning off the X-ray source). These reference images are used to compute the $K_D$ and $K_F$ matrices of parameters, which represent the empirical slopes of the $DF$ and $FF$ variations, respectively, as a function of the acquired projection index $i$. The index $i$ denotes the position of the $i$-th projection $CC_i$ within the sequence, allowing for its interpolated correction to yield the final image $IC_i$.

\begin{equation}
IC_i = \frac{CC_i-DF_0 \cdot (i \cdot K_{D} +1)}{FF_0 \cdot (i \cdot K_{F} +1)-DF_0 \cdot ( i \cdot K_{D} +1)}
\label{Ecu:corr}
\end{equation}

This empirical technique effectively mitigates image ghosting and the continuous increase in dark current as the sensor is irradiated during the tomography, as will be shown in the examples below in Fig.~\ref{fig:flat_field_corrected}.

\subsection{Application of conventional reconstruction algorithms}

Tomographic reconstruction from the projection dataset was performed using algorithms implemented in the \emph{Reconstruction Toolkit} (RTK) \cite{Rit_2014}. This toolkit includes several reconstruction methods, among them the FDK (Feldkamp-Davis-Kress) algorithm, which is the standard analytical filtered backprojection algorithm for cone-beam geometries \cite{Feldkamp:84}.

To accurately determine the center of rotation (CoR) of the projections, an optimization method based on maximizing the standard deviation of the reconstructed image was implemented \cite{walls2005correction}. Initially, a visual estimation of the CoR was obtained by comparing the first projection ($0^\circ$) with its corresponding $180^\circ$ pair. Around this estimated value, a range of candidate positions was evaluated by reconstructing a single cross-sectional slice for each one and calculating its standard deviation. For phase-contrast tomographies, the optimal CoR was fine-tuned to maximize this metric, as the narrow edge-enhancement fringes make the reconstruction extremely sensitive to misalignments. Conversely, for macroscopic objects scanned in absorption-contrast mode, the visual estimation of the CoR proved to be sufficient without further optimization.

After applying the reconstruction algorithms, the resulting 3D volume is typically visualized through cross-sections, enabling the observation of internal structures at different depths. In addition, to match the conventional color scheme used in tomography, we inverted the resulting image---background black, highest density in white---to make the images more familiar to the reader.

\section{Results}

This section presents two examples of high-spatial-resolution tomography acquired using the system described in the previous sections. 

First, we present the tomography of an integrated circuit in a surface-mount package. Since this device is composed of materials with high X-ray attenuation, it is suitable for analysis using conventional absorption-contrast tomography. The inspection of encapsulated integrated circuits is of particular interest for failure analysis within the microelectronics industry.

Second, the tomography of an insect is presented. Since this specimen is composed of materials with low X-ray absorption, it is particularly suitable for study using phase-contrast techniques. The analysis of insects and other biological samples is of significant interest in biological research and, potentially, in the analysis of clinical samples, such as biopsies for anatomical pathology. In both instances, tomographic reconstruction was performed using the FDK algorithm, selected for its favorable balance between result quality and computational efficiency \cite{10.1259/bjr.20180447}. To summarize the experimental parameters used for the acquisition of these examples, the most representative values are listed in Table~\ref{Tab:experiments}.

\begin{table}[!t]
\caption{Experimental parameters for the examples presented in the application of the tomographic technique.}
\label{Tab:experiments}
\centering
\begin{tabular*}{\columnwidth}{l@{\extracolsep{\fill}}c@{\extracolsep{\fill}}c}
\hline
Parameter & Example 1 & Example 2\\
\hline
Object & Integrated circuit & Insect \\
Object-to-sensor distance & 25\,mm & 190\,mm \\
Tube voltage & 35\,kV & 25\,kV \\
Tube current & $250\,\mu\text{A}$ & $100\,\mu\text{A}$ \\
Additional filter & Aluminum ($20\,\mu\text{m}$) & None \\
Angular step & 0.6\,$^{\circ}$ & 0.6\,$^{\circ}$ \\
Phase contrast & No & Yes \\
Field of view & $6.66 \times 5.32\,\text{mm}^2$ & $4.95 \times 4.00\,\text{mm}^2$ \\
\hline
\end{tabular*}
\end{table}

\begin{figure}[!t]
\centering\includegraphics[width=0.94\columnwidth]{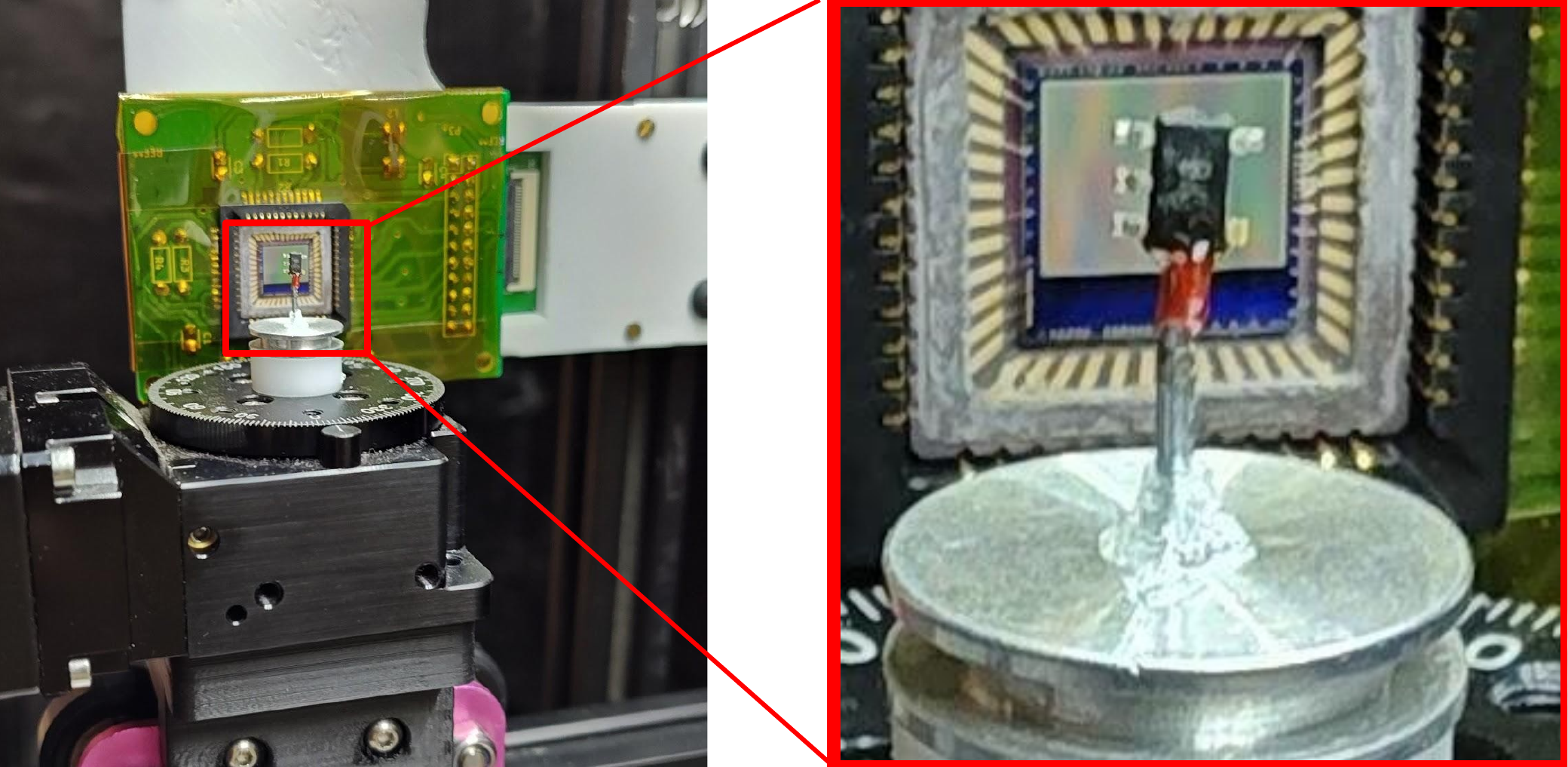}
\caption{The tomographed integrated circuit, placed 25\,mm from the sensor. The chip comprises a silicon die connected by gold wires to the metal terminals, and surrounded by a plastic package. 
}\label{Fig:chip_en_setup}
\end{figure}

\begin{figure}[!t]
\centering\includegraphics[width=\columnwidth]{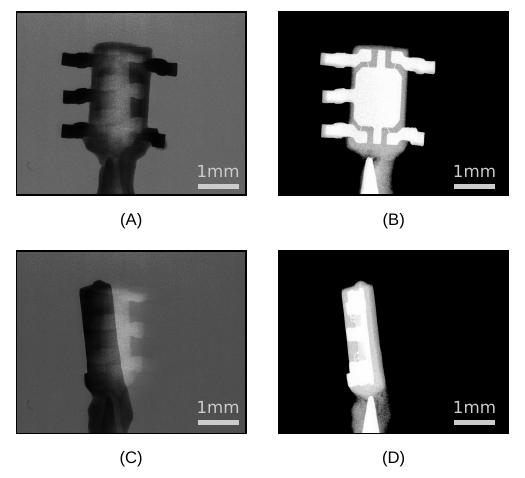}
\caption{Projections acquired by imaging the chip at two different angles (A, C) and results obtained after applying the defective pixel and interpolated flat-field corrections (B, D). Projections were acquired with a tube voltage of 35\,kV and a current of $250\,\mu\text{A}$.
}\label{fig:flat_field_corrected}
\end{figure}

\subsection{Conventional absorption-contrast tomography}
\label{sec:chip}

For this type of image acquisition, the object must be placed close to the sensor relative to the source-to-object distance. At this distance, phase-contrast fringes fall within the size of a pixel, and the acquired projections exhibit no observable phase contrast. Therefore, this geometry is suitable for analyzing high-attenuation samples.

The integrated circuit used as an example was an AD8615 operational amplifier in a surface-mount TSOT-23 package, with a size of 2.90\,mm$\times$2.80\,mm including the tips of the pins. It was positioned 25\,mm from the sensor, as shown in Fig.~\ref{Fig:chip_en_setup}, where the object magnification is negligible and the FOV is $6.66\,\text{mm}\times5.32\,\text{mm}$. In this case, 300 projections were acquired, with a step of $0.6^\circ$, covering a total range of $180^\circ$. The total acquisition time was approximately 4.24 hours, comprising 13020\,s for the projections and an additional 2250\,s for capturing the auxiliary Flat Frame and Dark Frame reference images.

To reduce beam hardening effects in the reconstruction, a $20\,\mu\text{m}$ aluminum foil was placed at the X-ray tube output to filter low-energy photons. To compensate for the attenuation induced by this filter, the tube was set to operate at a voltage of 35\,kV and a current of $250\,\mu\text{A}$. The raw averaged projections and the results of applying preprocessing to selected projections of the integrated circuit are shown in Fig.~\ref{fig:flat_field_corrected}.

\begin{figure}[!t]
\centering\includegraphics[width=\columnwidth]{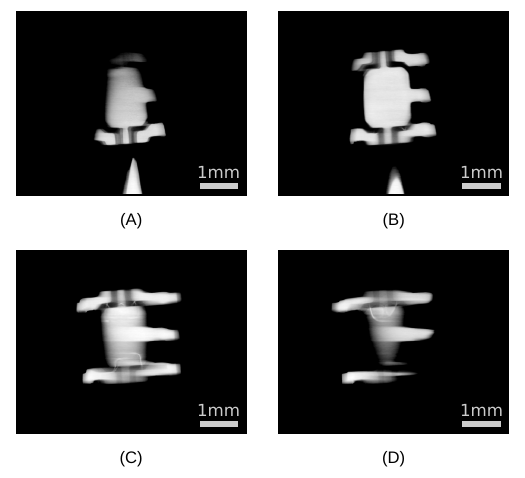}
\caption{Cross-sections of the reconstructed chip volume, obtained at $200\,\mu\text{m}$ depth intervals, revealing the internal metal components. Gold bonding wires used for circuit interconnection, which have an approximate diameter of $18\,\mu\text{m}$, are visible in image (C) and (D).
}\label{Fig:cross_section_chip}
\end{figure}

\begin{figure}[!t]
\centering\includegraphics[width=0.85\columnwidth]{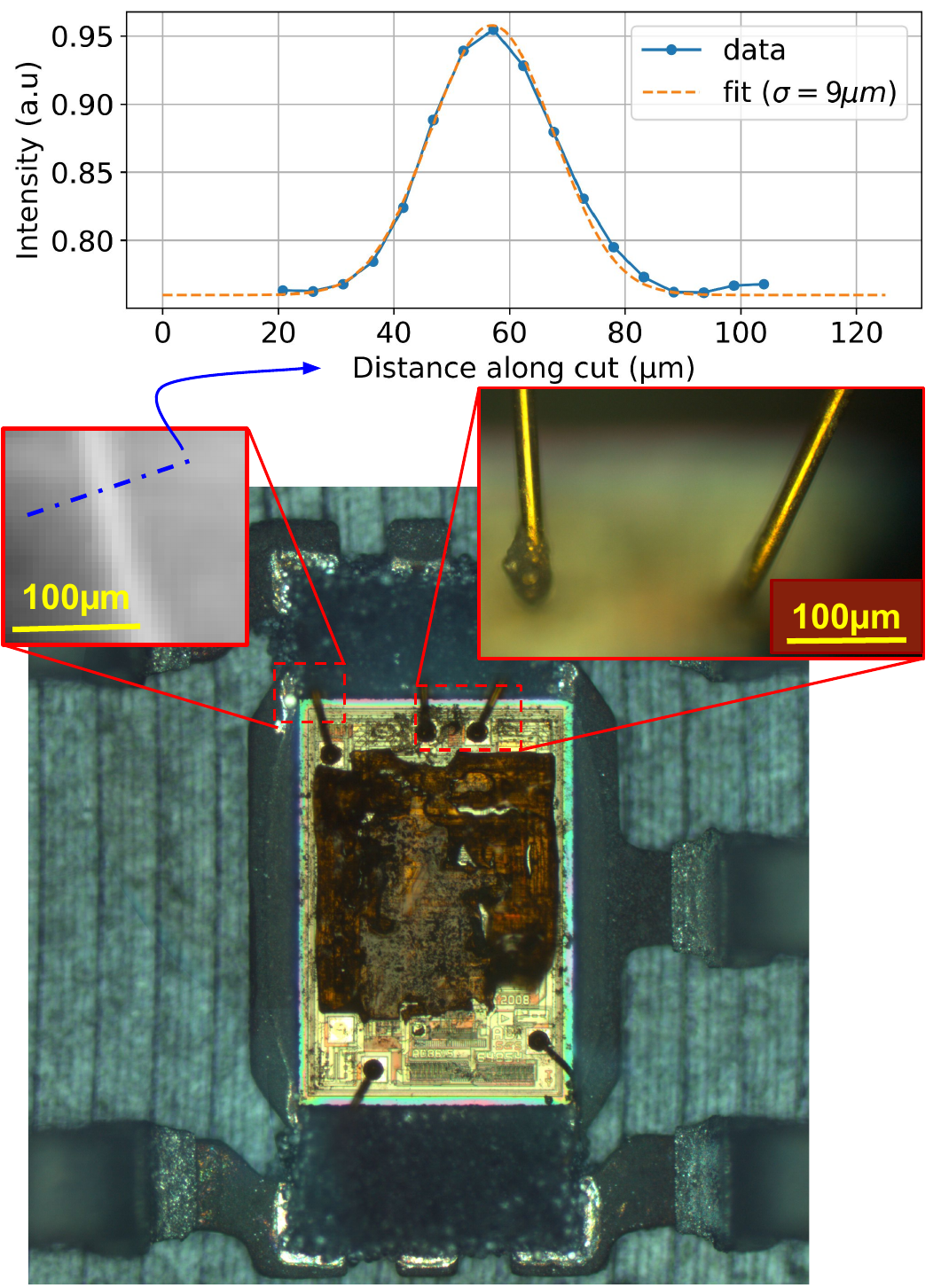}
\caption{Optical microscope photograph of the decapsulated AD8615 chip, showing the silicon die mounted on the lead frame. The image features a scale bar used to directly measure the approximate $18\,\mu\text{m}$ diameter of the gold wire bonds, providing a physical validation for the high spatial resolution achieved in the tomographic reconstructions.}
\label{Fig:chip_optical_photo}
\end{figure}

After applying the reconstruction algorithms on the sinograms, the results are presented in Fig.~\ref{Fig:cross_section_chip}, which displays cross-sections of the reconstructed volume. In these images, it is possible to appreciate the structure of the metallic chip terminals. Furthermore, cross-sections identified as (C) and (D) clearly show the gold bond wires. The $(18\pm4)\,\mu\text{m}$ diameter of these wires was verified via optical microscopy, as shown in Fig.~\ref{Fig:chip_optical_photo}. Two insets show a detailed optical image of the wires and the corresponding reconstructed portion. The visibility of such fine structures is of particular interest for failure analysis applications in the semiconductor industry and as a first step in decapsulating chips for single event effects testing with heavy ions.

To quantify the spatial resolution of the reconstruction, we analyzed the intensity along a cross section of one of the wires. The inset at the top of Fig.~\ref{Fig:chip_optical_photo} shows the intensity profile along the cut line crossing the wire (experimental points), together with a curve fitting the experimental data, obtained by convolving an $18\,\mu\text{m}$ pulse function with a Gaussian curve. The sigma required to fit the curve was $9\,\mu\text{m}$, corresponding to a full width at half maximum (FWHM) of $21\,\mu\text{m}$, which provides an estimate of the spatial resolution achievable with our reconstruction method.

\subsection{Phase-contrast tomography}
\label{sec_avispa}

Phase-contrast acquisition techniques enable the observation of objects that would otherwise be impossible to visualize using X-rays due to their low attenuation. Acquiring these images requires increasing the object-to-sensor distance, which simultaneously enables geometric magnification. However, since increasing this distance also increases geometric penumbra, the use of an X-ray tube with an effective focal spot comparable to the sensor's pixel size is justified.

\begin{figure}[!t]
\centering\includegraphics[width=\columnwidth]{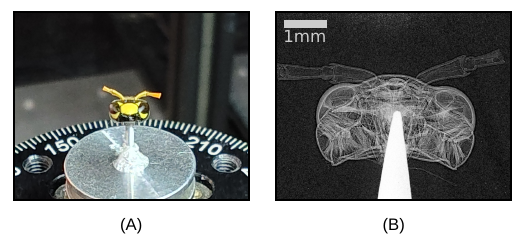}
\caption{Wasp head used as a sample for phase-contrast radiographic imaging, positioned 190\,mm from the sensor. Photograph of the mounted head (A). Corrected projection of the insect acquired from a frontal view (B).
}\label{Fig:abeja_en_setup}
\end{figure}

As an example, the head of a wasp (\textit{Vespula germanica}) was scanned, as shown in Fig.~\ref{Fig:abeja_en_setup}(A). The head width, measured across the widest point of the head, excluding the antennae, was 3.35\,mm. This specimen was placed at a distance of 190\,mm from the sensor to replicate the position defined in previous works \cite{MartinFase}. At this distance, the phase-contrast fringes are stretched out and fall across different pixels, becoming visible. In this position the object has a small magnification of $\approx$ 33\%, with a FOV of $4.95\,\text{mm}\times4.00\,\text{mm}$. For this acquisition, 600 projections were acquired, with a step of $0.6^\circ$, covering a total range of $360^\circ$ in a cone-beam geometry. Since the phase-contrast acquisition covered $360^\circ$ instead of $180^\circ$, twice as many projections were acquired, resulting in a total scan duration of approximately 7.86 hours (26040\,s for the projections and 2250\,s for the auxiliary files).

For phase-contrast acquisitions, a softer beam is desirable because the sensor response is generally higher at lower X-ray energies and the separation between interference fringes is larger, favoring the visibility of phase-contrast features. Therefore, the aluminum filter was omitted and the source was set to 25\,kV and $100\,\mu\text{A}$. The projections obtained after correction are displayed in Fig.~\ref{Fig:abeja_en_setup}(B).

\begin{figure}[!t]
\centering\includegraphics[width=\columnwidth]{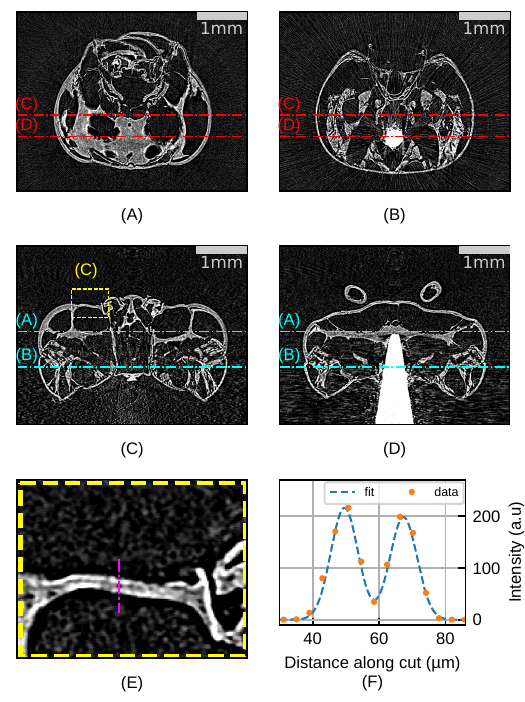}
\caption{Cross-sections of the reconstructed volume of the wasp head: (A) and (B) coronal cuts, and (C) and (D) sagittal cuts. The dashed lines indicate the position in which the orthogonal cuts were made. (E) Detail of the region presented in (A) with yellow dotted lines, and (F) an intensity plot along the pixels represented in (E), modeling two adjacent surfaces with two Gaussian curves.
}\label{Fig:cuts:wasp}
\end{figure}

The reconstruction results are presented in Fig.~\ref{Fig:cuts:wasp}, which displays cross-sections of the reconstructed volume. In this example, phase contrast enhances the sample edges, enabling improved visualization and reconstruction of a low-attenuation object that is difficult to analyze based only on conventional absorption techniques. This achievement is facilitated by the fine spatial sampling provided by CMOS image sensors, combined with the lateral coherence and minimal geometric penumbra of the microfocus X-ray tube.

To estimate the effective spatial resolution achieved in phase-contrast mode, we analyzed the intensity profile across two adjacent surfaces, as presented in Fig.~\ref{Fig:cuts:wasp}(F), where all distances are referred to the object plane. Unlike the chip example, where no two adjacent lines were available for a direct analysis of the system's line-resolving capability, the shapes present in the wasp's head provide more useful data for this purpose. The curve fitting the data was obtained as the sum of two Gaussian peaks separated by $17.8\,\mu\text{m}$, both requiring a sigma of $4.08\,\mu\text{m}$, corresponding to an effective FWHM of $9.6\,\mu\text{m}$, with the two surfaces clearly distinguishable. Given the $1.33\times$ geometric magnification used in this configuration, this corresponds to an effective FWHM of $\approx 12.8\,\mu\text{m}$ ($\approx 2.5\,$pixels) at the sensor plane, still sharper than the $21\,\mu\text{m}$ FWHM obtained for the absorption-contrast case. This improvement can be attributed to the edge-enhancement effect of phase contrast, which produces sharper edges in the reconstructed images in addition to the resolution gain from geometric magnification. Fitting similar edges in other parts of the images provides comparable results.

\begin{figure}[!t]
\centering\includegraphics[width=0.85\columnwidth]{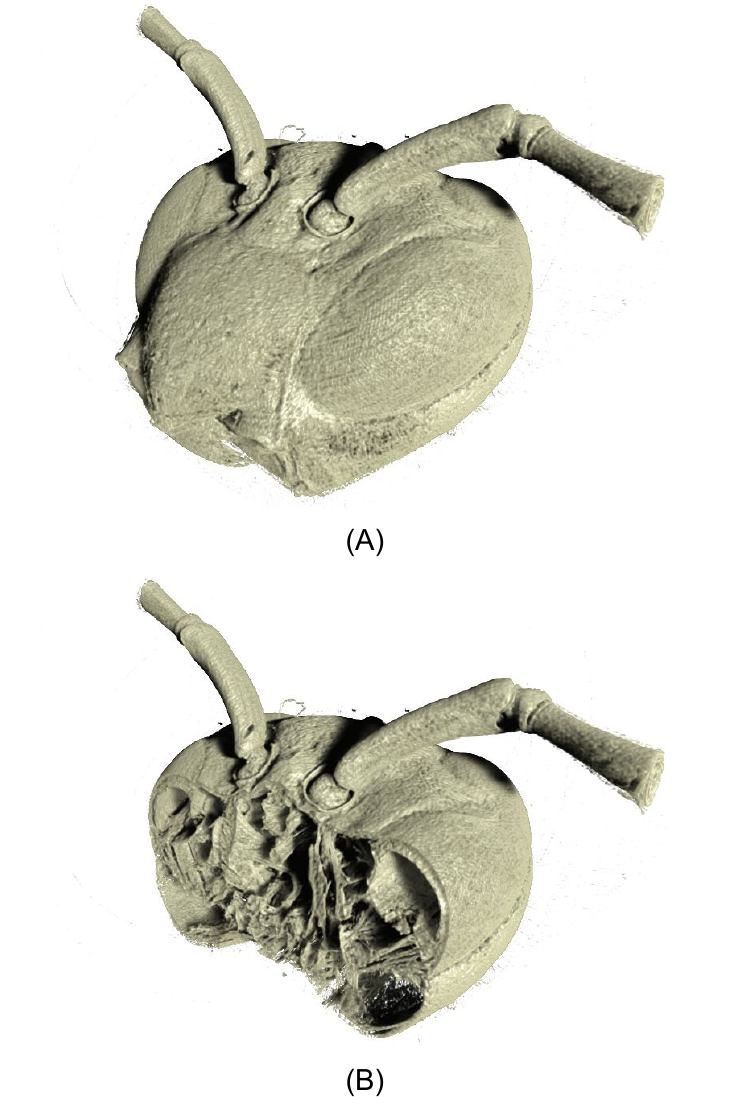}
\caption{Visualization of the reconstructed volume of a wasp head, generated using the 3D visualization tools in Dragonfly software. Full volume rendering (A). The same volume after applying a clipping plane to remove the frontal section of the head, revealing the different internal structures (B).
}\label{Fig:avispa3D}
\end{figure}

The acquisition and tomographic reconstruction workflows can be seamlessly integrated into open-source, Python-based pipelines using tools such as the Tomopari plugin for Napari \cite{Matotomo, napari_contributors_2019_3555620}. Subsequently, for advanced 3D rendering and visual inspection, the reconstructed volume can be explored from various orientations using software such as Dragonfly \cite{Dragonfly_20XX}. Figure~\ref{Fig:avispa3D}(A) displays a full volume rendering of the insect's head, while Fig.~\ref{Fig:avispa3D}(B) shows the same volume with a clipping plane applied to reveal the internal anatomical cavities.

\begin{figure}[!t]
\centering\includegraphics[width=\columnwidth]{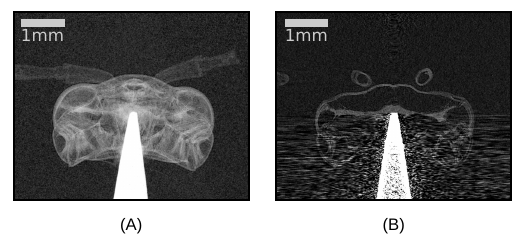}
\caption{Head projections acquired using conventional absorption contrast, with the sample positioned 25\,mm from the sensor. Frontal projection showing lower contrast than that obtained with phase-contrast projections (A). Cross-section of the reconstructed volume, showing higher noise levels in the rows aligned with the mounting needle (B).} \label{Fig:comp_ph_ac}
\end{figure}

To highlight the phase-contrast effect on the low-attenuation sample, we repeated the tomographic acquisition but placed the wasp head at only 25\,mm from the sensor. In this configuration, phase-contrast effects become negligible at this short distance, and the images are dominated by absorption contrast. Fig.~\ref{Fig:comp_ph_ac}(A) shows an example projection acquired in this configuration. The lower contrast makes it difficult or impossible to resolve certain details, particularly in the regions corresponding to the antennae.

The reduction in contrast affects both the visualization of the antennae and the accurate reconstruction of regions near the mounting needle used as mechanical support. This can be appreciated in Fig.~\ref{Fig:comp_ph_ac}(B), where the noise level is higher in the pixel rows corresponding to the mounting needle's position. This issue makes it impossible to segment the regions corresponding to the wasp's head by applying contrast adjustments or data binarization.

\color{black}

\section{Discussion and conclusions}
\color{black}

This study demonstrates that low-cost, COTS CMOS image sensors can serve as direct detectors for high-resolution tomography, supporting both absorption- and phase-contrast imaging without optical components or scintillators. The cone-beam setup uses a microfocus X-ray beam coherent enough to produce fringes resolvable by the CMOS pixel size.

Depending on magnification, voxel size in this work ranges from $3.9$ to $5.2\,\mu\text{m}$. Fitting the reconstructed bond wires gave a FWHM of $21\,\mu\text{m}$ ($\approx 4$ sensor pixels) in absorption-contrast mode, versus an effective FWHM of $9.6\,\mu\text{m}$ ($\approx 2.5$ pixels) at the object plane for adjacent surfaces separated by $\approx 17\,\mu\text{m}$ in phase-contrast mode. Using the larger FWHM value as a conservative criterion and assuming a Gaussian point-spread function, the system resolves $\approx 38\,\text{lp/mm}$ (at 10\% MTF).




Submicrometric spatial resolutions usually require synchrotron radiation \cite{albers2024high}: \cite{yakovlev2022wide} achieves a physical spatial resolution of $0.90\,\mu\text{m}$ (reconstructed at a $0.5\,\mu\text{m}$ voxel size) over a $5.0 \times 3.5\,\text{mm}^2$ field of view, and \cite{humbel2025synchrotron} reaches a physical resolution of $1.7\,\mu\text{m}$ (using a $0.65\,\mu\text{m}$ sampling grid) over a $1.3 \times 1.3\,\text{mm}^2$ area, extended to several millimeters for mouse-brain CT via stitching with a scientific CMOS detector; both are absorption-based. More complex synchrotron phase-contrast techniques, using Fresnel zone plates and high-resolution cameras, reach voxels of $15\,\text{nm}$ over much smaller fields of view \cite{wittl2026comparing}.

On the other hand, nano-CT systems based on special X-ray sources can also achieve resolutions of tens of nanometers, but at the cost of extremely complex and expensive emitters. For example, such resolution can be obtained by focusing a scanning electron beam onto a small (nanometer-scale) metallic anode and using geometric magnification \cite{wittl2026comparing}. Most of these systems rely on scintillators and complex optics to project the light generated by radiation onto an image sensor. These systems typically involve substantially higher costs than the setup presented here.

The use of much less expensive microfocus X-ray sources, as in this work, allows resolutions of a few micrometers. For instance, \cite{bartels2013phase} achieves $2\,\mu\text{m}$ resolution in propagation-based phase-contrast CT using a $7\,\mu\text{m}$ X-ray focal spot, projecting the images onto a scintillator read by a CCD with complex optical magnification, or alternatively using geometric magnification with a Medipix detector having a $55\,\mu\text{m}$ pixel pitch. Other works (see the list in \cite{navarrete2023x}) use grating-based interferometry or a modulator together with Medipix or flat-panel detectors under geometric magnification \cite{navarrete2023x}, but at the cost of very slow scans. Compared to these implementations, our setup implies a simple, straightforward, and less expensive configuration, based on a microfocus X-ray source that is much more available than synchrotrons and nanofocus X-ray beams, a simple arrangement without X-ray or visible-light optics, and a CMOS image sensor as an affordable direct X-ray detector. The key advantage is that the pixel size and spatial resolution of the COTS CMOS sensor are well matched to the microfocus source's focal spot and geometric magnification, providing adequate spatial resolution for high-contrast tomographic reconstructions without over-sampling or under-sampling the projection data. The cost of the whole setup could even be lowered using conventional X-ray sources without micrometric focal spots, if only absorption-contrast images are needed.

The use of CMOS image sensors for such long irradiations requires special techniques to handle sensor aging from total ionizing dose (TID) effects, which increase both dark and flat fields. We previously used a standard flat-field correction with single correction images \cite{corzi2024enhanced}, but prolonged irradiations cause large variations in the correction fields, requiring the dynamic flat-field correction proposed here. This modified technique compensates for progressive sensor degradation from accumulated dose, enabling prolonged acquisitions; the simple interpolation proved sufficient, without the more complex corrections used in \cite{jailin2017use}.

The combination of limited microfocus X-ray power, the long object-to-sensor distances needed to observe interference fringes \cite{MartinFase}, the limited detection efficiency \cite{corzi2024enhanced}, and the available FOV implies that CT acquisition takes a few hours. This limits its applicability to untreated hydrated samples and living, moving insects, for which faster detectors and higher-flux (e.g. synchrotron) sources are required \cite{albers2024high, TANG2026104073}.

With these considerations, we obtained CT in two modes: absorption contrast, resolving $18\,\mu\text{m}$-diameter bond wires in integrated circuits (FWHM $\approx 21\,\mu\text{m}$), and propagation-based phase contrast, resolving adjacent surfaces $\approx 17\,\mu\text{m}$ apart in biological samples (effective FWHM $=9.6\,\mu\text{m}$ at the object plane), revealing soft-tissue boundaries that are difficult to visualize using absorption-contrast imaging in the same experimental configuration. These results confirm that consumer-grade CMOS sensors are an accessible, powerful alternative for high-resolution microtomography.

\section*{Acknowledgment}
\addcontentsline{toc}{section}{Acknowledgment}
\scriptsize

The authors would like to thank Julio Marin for providing the rotation stage, and the Radiation Protection Division (CNEA) assistance with radioprotection. AI tools (Grammarly, DeepSeek, Gemini, Claude) were used only as coding assistants and for English grammar correction by the non-native authors. No technical or scientific content was generated by AI.

\normalsize

\bibliographystyle{IEEEtran}
\bibliography{mibib}

\end{document}